\documentclass[12pt,epsfig]{article}
\usepackage{graphicx}
\usepackage{amssymb}
\begin{document}

\baselineskip=18.6pt plus 0.2pt minus 0.1pt

 \def\be{\begin{equation}}
  \def\ee{\end{equation}}
  \def\bea{\begin{eqnarray}}
  \def\eea{\end{eqnarray}}
  \def\nn{\nonumber\\ }
\newcommand{\nc}{\newcommand}
\nc{\bib}{\bibitem} \nc{\cp}{\C{\bf P}} \nc{\la}{\lambda}
\nc{\C}{\mbox{\hspace{1.24mm}\rule{0.2mm}{2.5mm}\hspace{-2.7mm} C}}
\nc{\R}{\mbox{\hspace{.04mm}\rule{0.2mm}{2.8mm}\hspace{-1.5mm} R}}

\begin{titlepage}
\title{
\begin{flushright}
 {\normalsize \small
GNPHE/0912 }
 \\[1cm]
 \mbox{}
\end{flushright}
{\bf Holonomy Groups  Coming From  F-Theory  Compactification  }
\author{
Adil Belhaj$^{1,2,3}$\thanks{\tt{belhaj@unizar.es}}, Luis J.
Boya$^2$\thanks{\tt{luisjo@unizar.es}},
Antonio  Segu\'{\i}$^2$\thanks{\tt{segui@unizar.es}}\\[.3cm]
{ \small $^1$ Centre National de l'Energie, des Sciences et des Techniques Nucl\'eaires,  Rabat, Morocco} \\
{ \small
$^2$ Departamento de F\'{\i}sica Te\'orica, Universidad de Zaragoza,
E-50009-Zaragoza, Spain}
\\
{\small $^3$ Groupement National de Physique des Hautes Energies, Si\`{e}ge focal: 
FSR, Rabat,  Morocco }\\
 } } \maketitle
\thispagestyle{empty}
\begin{abstract}
We  study holonomy groups coming from F-theory   compactifications. We focus mainly on  $SO(8)$  as $12-4=8$ and subgroups $SU(4)$, $Spin(7)$, 
 $G_2$ and  $SU(3)$ suitable for  descent from F-theory,     M-theory and  Superstring theories. We consider the  relation  of these groups with
 the octonions, which  is striking and reinforces their role in higher dimensions and dualities.  These holonomy groups are related 
 in various mathematical forms,   which we  exhibit.
\end{abstract}
{\tt  PACS}: 02.20Hj, 14.80Rt, 11.25-w\\
{\tt  KEYWORDS}:  Superstrings, M-theory, F-theory,  Octonions,  Holonomy Groups.
\end{titlepage}
\tableofcontents
\newpage
\newpage
\section{Why extra dimenions}

Once one includes strings and/or other extended objects, extra dimensions became unavoidable: for example, particles dualize in four dimensions, like electrons and magnetic monopoles, but strings dualize in six, and membranes in eight. On the other hand, e.g. string fields like to share supersymmetric partners, so Susy also is more cogent in higher dimensions. Hence the necessity of dimensional reduction, introducing tiny compact spaces, as we see only four extended dimensions \cite{vafa}.

In this paper we  discuss  the geometry  of these compact spaces from the point of view of holonomy groups and string dualities; we focus mainly on the mathematical side of the new constructions. The preferred extra freedoms are needed for a total of ten, eleven or twelve dimensions: superstrings live in ten (whereas the bosonic string needs 26), but M-Theory (Witten, 1995) prefers 11 \cite{witten,townsend,PT,Duff,AW,adil1,adil2,adil4,adil3} , and F-Theory (Vafa, 1996, 2008) uses 12 
\cite{vafaft,Boya2,Boya1,BHV1,vafa08,ABBS,ABBMS}.

Over the past few years, there has been an increasing interest in studying duality in relation to supersymmetry and compactified manifolds \cite{GS1,sen,Oo,As,Candelas,witten1,vafa}; by duality we first understand the naive concept that a field strength $F$ of dimension $d$  in a manifold of dimension $D$ dualizes to another field strength $F^*$ of dimension $D-d$. For  example, strings couple to potential 2-forms $B_{\mu\nu}$, hence to a 3d field strength, and in $D=10$ dimensions the dual of a fundamental string is a 5-brane. More geometric is the duality between some Calabi-Yau (CY) spaces, discovered by Candelas {\it et al.}   in their search for interesting  manifolds suitable for heterotic string compactification, called mirror symmetry \cite{Candelas1}. The most important consequence of the string duality was found by Witten: namely, the five viable superstring theories are special limits in the moduli space of the same theory, called M-Theory \cite{witten}. M-Theory was considered for about ten years the best candidate for the unification of microscopic forces and also with gravity, as the low energy limit of this theory describes the well-known 11-dim supergravity \cite{CJS}. However,
 M-Theory never illuminated anyone of the genuine  features of the microphysical world: neither gauge groups, nor particle spectrum nor even the number of distinct forces were selected by M-Theory. So lately C. Vafa has resurrected a theory of his, of 1996 (F-Theory, which lives in 12 dimensions \cite{vafaft}) with the assumption that perhaps decoupling gravity the new F-Theory, in adequate compactification, can account for some of the features of the standard model \cite{BHV1,vafa08}. In this paper we consider the geometric peculiarities of these dimension differences, for objects living in 10, 11 or 12 dimensions.

Within strings supersymmetry is mandatory, lest we want to contemplate 26 dimensions; but we can only tolerate ${\cal N}=1$ Susy in our mundane, $4d$  space. Parity violation is a conspicuous feature of our world, but with ${\cal N}>1$ Susy, chiral partners are in the same multiplet; so parity is conserved. The preservation of this feature puts stringent conditions on the compactifying manifold with $10-4=6$ dimensions, and corresponding results obtains from descent $11\to 4$ (M-Theory) and $12\to 4$ (F-Theory) 
\cite{vafaft,BHV1,vafa08}.

In particular, we need manifolds with $SU(3)$ holonomy groups for the heterotic string case \cite{Candelas}, essentially because $4=3+1$ in the descent $Spin(6)\cong SU(4)$  to $SU(3)$. The possible manifolds are signalled $CY_3$. In M-Theory the chain is: descent $Spin(7)$  to $G_2$, because 8=7+1 again. The restrictions on F-Theory depends on the signature (one or two times) and will be dealt with later. The further break from the gauge group (e.g. $E_8\times E_8$  in the case of heterotic exceptional superstring) to more realistic Sandard  Model  groups, like $E_8\times E_6$ , and further $E_6 \to SU(5)\to SU(3)_C\times SU(2)_L\times U(1)_Y$ requires additional mechanisms, of course, see e.g. 
\cite{witten1}. More recently, the descent from 12 to 4 in F-Theory is managed in two or three steps, see below.

As a whole, there are many ways to get four-dimensional models using different compactifications as intermediates. There are relations between several of these constructs due to special dualities which appear in the process. As an example, we point out here different equivalences in seven dimensions giving rise to a web of dualities (with F and M for F-theory resp. M-Theory)  \cite{vafa}
\begin{equation}
F/K3 \times S^1\sim het/T^3\sim  IIB/S^2\times S^1\sim M/K3\sim IIA/S^2\times S^1.
\end{equation}

This can be pursued, of course, to lower dimensions. The main focus of this paper is the study of these compactifications down to four dimensions within the perspective of dualities. We shall focus mainly in four  groups, $SU(4)$, $Spin(7)$,   $G_2$ and  $SU(3)$. All these can be seen as subgroups of $SO(8)$, the maximal  compact group  of the 
F-Theory compactification  down to four dimensions ($12 \to 4$):  This can be obtained by breaking the space-time symetry SO(1,11) down to the subgroup $SO(1,3)\times SO(8)$. We shall not be much concerned with the manifolds themselves.

On the other hand, the relation of the above groups with the octonion division algebra should be evident \cite{BBS}, as e.g. $Spin(7)$ acts in the 7-sphere of unit octonions, $G_2$ in the 6-sphere of unit imaginary octonions, and $SU(3)$ on the equator of the later; we devote some space in the Appendix to study in detail such connections.  We shall exhibit also  in section 4 some exact  triple sequences, which relate the precise mathematical relations between these holonomy  groups.

The organization of the paper is as follows. In Sect. 2 we recall the classification of special holonomy manifolds by Berger (1955). In Sect. 3 we review different ways of constructing four-dimensional models with minimal number of supercharges from higher dimensional supersymmetric theories.  Sect. 4 deals with F-theory and its  relations  with  holonomy groups, and exhibits also the exact sequences mentioned above. The  Appendix elaborates on generalities over the octonion division algebra.

\section{ Manifolds with  special holonomy}
\subsection{About  holonomy groups}
 The study of the supersymmetric theories  and extended  objects  involves  the prediction of extra dimensions of space-time. 
However, as we see only   $4= (1+3) $ dimensions, some mechanism has to be advocated to
 prevent  the extra size  of the space to be visible. The compactification is the most
 accepted ingredient, namely making the extra  dimensions too small to be observable. In the original
 Kaluza-Klein type of theories ( ca. 1920), the observable  gauge group  in 4 dimensions came from
the {\it isometry} groups of the compactifying space (this is why
the $U(1)$  gauge group
  of electromagnetism came from the compactification of the fifth  dimension on a circle).  But when supersymmetry is present,
it was realized in the early 80s  that  the {\it holonomy} groups of
intermediate spaces  respond of the number of supercharges surviving
in four  dimensions \cite{awada83}.

In this section  we review  briefly the classification of  special holonomy groups and manifolds  in a form suitable for all later physical
 consideration.  Note that
the books of Joyce \cite{Joyce,Joyce2}  are  the best modern references  for
this subject.
 Let  $\cal M$ be any  $n$ dimensional differentiable manifold. The structure group of the tangent bundle
is a subgroup of the general linear group, $GL(n,R)$. Now the
maximal compact group of the linear group is $O(n)$. So the quotient
homogenous space $ \frac{GL(n,R)} {O(n)}$ is a contractible space;
hence, a manifold admits always a Riemannian metric $g$, whose  tangent
structure  group  is (a subgroup of) the orthogonal group.  The isometry
group $Isom(\cal M)$  is the set of diffeomorphisms  $\sigma$ leaving
$g$ invariant. For spheres we have $Isom(S^n)=O(n+1)$; for torii  $Isom(T^n)=U(1)^n$.

For an arbitrary $n$ dimensional Riemannian manifold $\cal M$, the
structure group of the tangent  bundle is, as said, a subgroup of
$O(n)$. Carrying a orthonormal frame $\epsilon$ of $n$ vectors in a
point $P$ through  a closed loop $\gamma$ in the manifold,  by parallel transport back to $P$,
\begin{equation} 
\gamma: P \to P'\to P
\end{equation}
it becomes another frame $\epsilon'=o \cdot \epsilon $  which is  shifted by certain  element $o$ of $O(n)$.
This is the {\it holonomy element} of the loop. All elements of all possible loops on the manifold
 from $P$ to $P$  make up
the holonomy group of the  manifold $Hol_P(\cal M)$,  which is  always a subgroup of $O(n)$, and is easily
 seen to be independent of the point $P$ for an arcwise-connected manifold.
 A generic Riemannian manifold  would have holonomy $O(n)$, or $SO(n)$ if it is orientable, whereas the isometry
 group  is just the identity generically; in a way isometry and holonomy are complementary.

For any vector bundle with a connection, the structure group reduces to the holonomy group (reduction theorem).
The corresponding Lie algebra of the holonomy group is generated by the curvature of the connection
(the Ambrose-Singer theorem)\cite{KN63}.

 \subsection{ Types of special holonomy manifolds}
The classification
of special  holonomy groups was  carried by M. Berger in 1955. If
the manifold is irreducible, $ Hol ({\cal M})$ should lie in $ O(n)$.
Its Lie algebra, as we said, is generated by the curvature. For the
irreducible non symmetric case, there are three double series of classes of manifolds 
 corresponding to the numbers $\mathbf{R},\mathbf{C}$ and $\mathbf{H}$, and two isolated cases
  related to the octonion numbers $\mathbf{O}$.  For  each number domain there are the  generic
 case and the unimodular subgroup restriction.  The list practically  coincides with the list of groups with transitive action on spheres.\\
 The classification  of holonomy groups is given in the following Table \cite{Boya1}
\begin{center}
\begin{tabular}{lll}
Numbers & Group & Unimodular Form \\
\hline \\
$\mathbf{R}$ & $O(n)$ & $SO(n)$ \\
& generic case & orientable, $w_{1}=0$ \\
&  &  \\
$\mathbf{C}
$ & $U(n)$ & $SU(n)$ \\
& K\"{a}hler, $d\omega =0$ & Calabi-Yau, $c_{1}=0$ \\
&  &  \\
$
{\mathbf{H}}
$ & $Sp(1) \times/_2 Sp(n)$ & $Sp(n)$ \\
& Quaternionic &  Hyperk\"{a}hler \\
&  &  \\
$\mathbf{O}$ & $Spin(7)$ in $8d$ spaces & $G_{2}$ in $7d$ spaces \\
& $Oct(1)$ & Aut$\mathbf{(O)}$%
\end{tabular}
\end{center}

Some explanations are in order.  An arbitrary $n$-dimensional Riemannian manifold
$ \cal{M}$  has $O(n)$  as the maximal holonomy group. The obstruction
   to   orientability is measured  by the first  Stiefel-Witney class of the tangent bundle,
   $w_1({\cal M})\equiv w_1(T{\cal M}) \in H^1( {\cal M} ,Z_2)$.\\
A  $n$-dimensional complex K\"{a}hler  manifold parameterized  by $z_i, i=1,...,n$  has
 a closed  regular real K\"{a}hler  two-form  $\omega$ given in a local chart by
\begin{equation}
\omega=iw_{i{\bar j}}dz_i\wedge d{\bar z_{\bar j}}, \quad d\omega=0,
\end{equation}
where $w_{i{\bar j}}$ can be expressed  as 
\begin{equation}
w_{i{\bar j}}=\frac{\partial ^2K}{\partial z_i \partial {\bar z_{\bar j}}},
\end{equation}
and where $K$ is a locally defined function   called the K\"{a}hler  potential. The 
 holonomy group of  such a  geometry  is $U(n)$.  Now as $ \frac{U(n)}{SU(n)}=U(1)$, we have the diagram

\begin{equation}
\begin{tabular}{lllll}
$\;$$\;$SU(n) &  &  &  &  \\
$\quad$ $\;$ $\downarrow $ &  &  &  &  \\
$\;$ U(n) & $\longrightarrow $ & B & $\longrightarrow $ & M \\
det $\downarrow $ &  &  &  &  \\
$\;$$\;$U(1) & $\longrightarrow $ & B%
\'{}
& $\longrightarrow $ & M%
\\
$\quad$ $\;$  &  &  &  & 
\end{tabular}%
\end{equation}
where   the middle line  is the   frame bundle: $B$ is  the principal bundle of  unitary frames.
 The last bundle  is mapped  to an element of $H^2({\cal M}, Z)$; hence, the determinant map defines the
  first Chern class of the bundle as $c_1({\cal M}) \in H^2({\cal M},Z)$.
It turns out then  that when $c_1=0$, the K\"{a}hler manifold becomes a CY manifold  with
$SU(n)$  holonomy group and it is Ricci flat, $Ric=0$; this is because (\cite{Joyce}, p. 98), the Ricci tensor is equivalent to a 2-form proportional to $c_1$, which is  zero in the CY case. Note that an one-dimensional CY  manifold  is
nothing but a  torus $T^2$, as $SU(1)=1$. So  its Hodge diamond  is given by
\def\m#1{\makebox[10pt]{$#1$}}
\begin{equation}
  {\arraycolsep=2pt
  \begin{array}{*{5}{c}}
    &&\m{h^{0,0}}&& \\ &\m{h^{1,0}}&&\m{h^{0,1}}& \\
    &&\m{h^{1,1}}&&
  \end{array}} \;=\;
  {\arraycolsep=2pt
  \begin{array}{*{5}{c}}
    &&\m1&& \\ &\m1&&\m1& \\ &&\m{1}&&
  \end{array}}
\end{equation}
The second example of  CY  geometries  is the  K3 complex surface with  $SU(2)$ as holonomy group.
Its  Hodge diammond reads
\def\m#1{\makebox[10pt]{$#1$}}
\begin{equation}
  {\arraycolsep=2pt
  \begin{array}{*{5}{c}}
    &&\m{h^{0,0}}&& \\ &\m{h^{1,0}}&&\m{h^{0,1}}& \\
    \m{h^{2,0}}&&\m{h^{1,1}}&&\m{h^{0,2}} \\
    &\m{h^{2,1}}&&\m{h^{1,2}}& \\ &&\m{h^{2,2}}&&
  \end{array}} \;=\;
  {\arraycolsep=2pt
  \begin{array}{*{5}{c}}
    &&\m1&& \\ &\m0&&\m0& \\ \m1&&\m{20}&&\m{1.} \\
    &\m0&&\m0& \\ &&\m1&&
  \end{array}}
\end{equation}
Since $SU(2)=Sp(1)$, the K3  surface is also a hyperk\"{a}hler  manifold:
 Notice that hyperk\"{a}hler manifolds are also Calabi-Yau, but the quaternionic
 manifolds in general are not. Quaternionic manifolds have for holonomy
 $Sp(1) \times Sp(n)/Z_2$, abreviated as $Sp(1) \times/_2 Sp(n)$ in the Table.

Finally, the  two cases related to the octonions are  $G_2$, that is the octonion automorphism group  and $Spin(7)$.
The former is well known  and we shall elaborate on it later; as for the $"Oct(1)"$
label for $Spin(7)$, this  will also be clarified  in the Appendix.

Note that, in general,   a manifold  with a specific holonomy  group $ Hol( {\cal M})=G$
 implies  the manifold carries an additional structure, preserved by the group $G$. For
 example, an orientable  manifold, with holonomy  within $SO(n)$,   has an invariant volume element, indeed
$SO(n)=O(n)\bigcap SL(n,R)$.  A  K\"{a}hler  manifold, with holonomy inside $U(n)$
has an invariant closed 2-form as $U(n)=O(2n)\bigcap Sp(n)$; a $SU(n)$  holonomy manifold carries  a holomophic volume.   A manifold with $G_2$ holonomy
 will carry an invariant $3$-form, etc.

\section{  Physical compactification scenarios}
As we know  superstring  theory  lives in ten dimensions
\cite{GS1}; down to four dimensions we want only ${\cal N} =1$,
i.e. four supercharges, as to allow for parity violation. We know
that compactification on  a $SU(n)$-holonomy manifold would reduce
the supercharges by a factor of $1/2^{n-1}$, so $SU(3)$-holonomy
(i.e., a CY$_3$) would be just right to descend from
the heterotic string ($16$ supercharges) to a four  dimensional model with only four supercharges.  Indeed, the search for
CY$_3$ manifolds was a big industry in the $80$s \cite{Candelas}.
This choice is also natural, as  $SU(3) \subset SU(4)=Spin(6)
\rightarrow SO(6)$, and obviously as  $4=3+1$,  $SU(3)$   leaves one surviving
spinor.

In M-Theory  living  $11$ dimensions,  the candidate compactifying manifold would be one with $G_2$ holonomy group \cite{PT}:
now the inclusions are $G_2 \subset SO(7) \leftarrow Spin(7)$, and $8=1+7$, as $2^{(7-1)/2}=8$, type real.
$G_2$-holonomy manifolds  which are also Ricci flat,  see again (Joyce, \cite{Joyce}, p.244)  and would preserve $1/2^3$ supercharges, and in $11$d there are $2^{(11-1)/2}=32$,
type real again as $10-1=9 \equiv 1$ mod $8$.

We can also consider eight-dimensional  compactifying manifolds in at least two contexts: $1$) Descend $11 \rightarrow 3$ just
for illustrative purposes, and $2$) F-Theory with metric ($1,11$); the original sugestion of Vafa was $12=(2,10)$,
see \cite{vafaft, Boya2}. Here the manifolds of choice would be either CY$_4$, that is, $SU(4)$-holonomy manifolds,
preserving $4$ supercharges out of $32$ (which is what we want), or $Spin(7)$, the last of the exceptional
holonomy groups; $Spin(7)$ does the job as it has an irreducible $8$d representation, same as $Spin(8)$ and
$Spin(7) \subset Spin(8)$. The following Table sums up the situation:

 \begin{center}
\begin{tabular}{lll}
Theory & \quad Dim Change &  \qquad Holonomy \\
\hline \\
Heter. String & $\quad 10d\longrightarrow 4d$ & $\qquad SU(3)$ (CY$_3$ manifold, Ricci flat) \\
 & &  \\
M-theory  & $\quad 11d\longrightarrow 4d$ & $\qquad G_2$ (Ricci flat) \\& &  \\
M-theory  & $\quad 11d\longrightarrow 3d$ & $\qquad Spin(7)$ (Ricci flat) \\& &  \\
F-theory $(1,11)$  & $\quad 12d\longrightarrow 4d$ & $\qquad Spin(7), SU(4) ($CY$_4)$ \\& &  \\
F-theory $(2,10)$ & $\quad 12d\longrightarrow 4d$ & \qquad Indefinite form of $Spin(7)$ or $Spin(8)$. %
\end{tabular}
\end{center}
\bigskip

We note that if we consider the conventional F-Theory with signature $(2,10)$ it is necessary to compactify in
 a manifold with signature $(1,7)$.

\section{ Connections  between holonomy groups from F-theory }
 We consider now   in the spirit of Vafa's  new F-theory compactifications \cite{BHV1,vafa08} relations between special holonomy groups.

\subsection{ Holonomy groups from  F-theory compactification}
In F-theory compactification, with one time,  the four dimensional  models can be obtained by breaking the space-time symmetry $SO(1,11)$ down to the following subgroup
\begin{equation}
 SO(1,11)\to  SO(1,3)\times SO(8)
\end{equation}
where $SO(8)$ is  the maximal holonomy group of an  eight-dimensional  intermediate  manifod $X_8$.  All above discussed special holonomy group   can be related to $SO(8)$ symmetry.
As we know the Dynkin   diagram for $SO(8)$ is given by 

\begin{figure}[tbph]
\centering
\begin{center}
\hspace{1.5cm} \includegraphics[width=6cm]{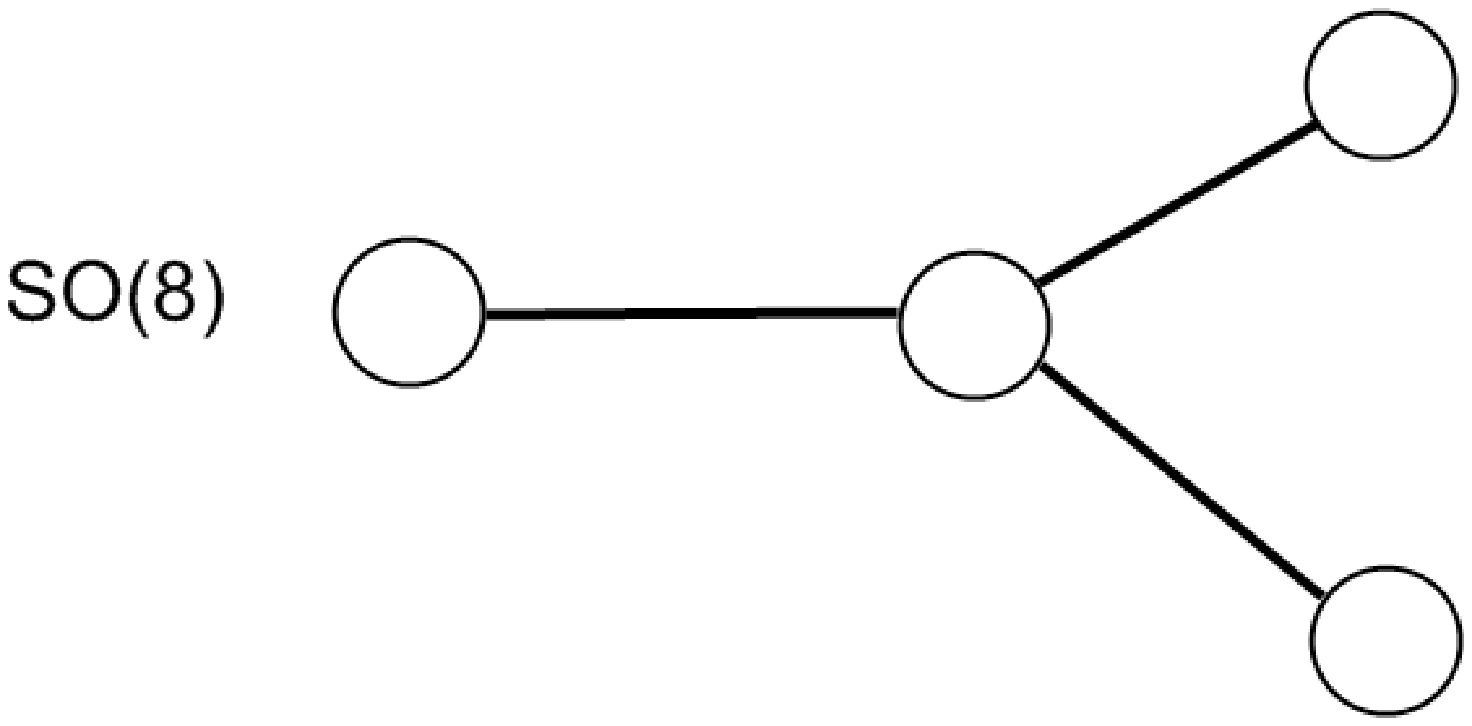}
\end{center}
\end{figure}
It is remarakble how the different  special honolomy groups come  from this  Dynkin diagram.          
 In particular, $SU(4)$   and SU(3)  can be obtained by  deleting one and two nodes respectively
\vspace{1.5cm}
\begin{figure}[tbph]
\centering
\begin{center}
\hspace{1.5cm} \includegraphics[width=6.5cm]{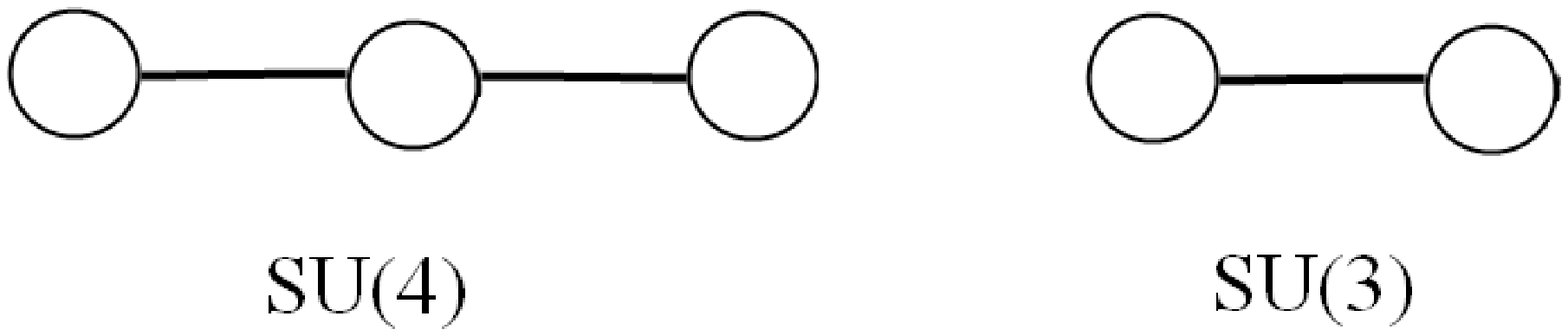}
\end{center}
\end{figure}

The Dynkin diagram of  $SO(8)$ shows triality, the free permutations of the three external nodes. Identifying these nodes, we get the $G_2$  as  the fix point set  of  $S_3$ group. The $ Spin(7)$ group can be obtained  by identifying   the right  spin  nodes:
\vspace{1cm}
\begin{figure}[tbph]
\centering
\begin{center}
\hspace{1.5cm} \includegraphics[width=8cm]{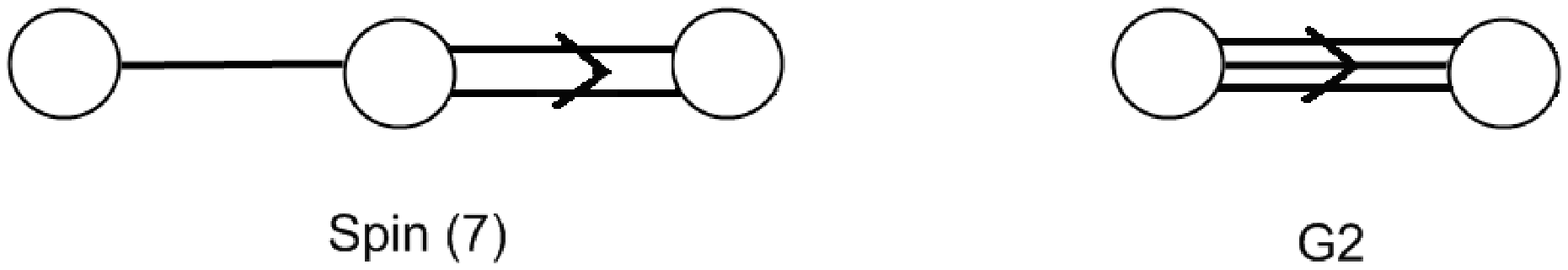}
\end{center}
\end{figure}

As we  see,  all special holonomy groups can be related to the maximal holonomy of the  F-theory compactifications down to  four dimensions.

\subsection{ Relations between different special  holonomy groups}

Strings,  M and F-theories are related by different sorts of dualities and  compactifications. As a consequence we expect that also the different holonomy groups used in  various
  compactifications should be connected. In what follows  we  address this question  using   exact sequences and commutative diagrams
for these groups.  To start,  we  note the following. If $H \subset G$  with (left)-coset space $X$,
 we write $H \to G  \to  X$ for  $ G/H =  X$; when $H$ is normal, $X$ becomes the quotient group.
   Some of next  diagrams   have been already  given  in \cite{Boya3}.\\
The  first diagram   that we present here  comes form the inclusion  of the exceptional holonomy group
$ G_2 \subset Spin(7)$. The later acts transitively in all units in $\mathbf{O}$ (octonions of norm one, forming $S^7$), whereas $G_2= Aut(\mathbf{O})$  obviously leaves $1$
  invariant (the real part of the octonion).
 So  the main cross of the diagram  takes the following form 
\bigskip

\begin{equation}
\begin{tabular}{lllll}
&  & Spin(6) &  &  \\
&  & $\quad \downarrow $ &  &  \\
$G_{2}$ & $\longrightarrow $ & Spin(7) & $\longrightarrow $ & $S^{7}$ \\
&  & $\quad \downarrow $ &  &  \\
&  & $\quad S^{6}$ &  &
\end{tabular}%
\end{equation}
where the vertical line is elemental\footnote{The spin groups acting  on  the natural spheres via the $SO$
 (covered) groups, $Spin(n)/Z_2=SO(n)$.}. With the $A_3=D_3$ isomorphism $Spin(6)=SU(4)$  and the fact that
  $ SU(3)\subset G_2\bigcap (Spin(6) =SU(4))$, we can complete the previous cross.  The result is given by the diagram
\begin{equation}
\label{seq}
\begin{tabular}{llllll}
$SU(3)$ & $\longrightarrow $ & $SU(4)$ & $=Spin(6)$ & $\longrightarrow $ & $%
S^{7}$ \\
$\quad\downarrow $ &  &  & $\downarrow $ &  & $\parallel $ \\
$\quad G_{2}$ & $\longrightarrow $ &  & $Spin(7)$ & $\longrightarrow $ & $S^{7}$
\\
$\quad\downarrow $ &  &  & $\downarrow $ &  &  \\
$\quad S^{6}$ & === &  & $S^{6}$ &  &
\end{tabular}.
\end{equation}
From this picture   we  can see in particular the octonionic nature
of $SU(3)$. It is a group of
 automorphism of octonions, fixing  the product, say  $(ij)k$. There is a suspicion, still conjectural,
  that this is the reason why the gauge group of the strong forces  is  $SU(3)$ color.\\
To get  the second diagram,  we start by another  obvious cross, since   $Spin(7)$  is the (universal)
 double cover of  $SO(7)$.  In this way, we have the following diagram
\begin{equation}
\begin{tabular}{lllll}
&  & $Z_{2}$ &  &  \\
&  & $\downarrow $ &  &  \\
$G_{2}$ & $\longrightarrow $ & $Spin(7)$ & $\longrightarrow $ & $S^{7}$ \\
&  & $\downarrow $ &  &  \\
&  & $SO(7)$ &  &
\end{tabular}.
\end{equation}
It is  known  that $G_2$ does not  have a   centre, so $Z_2=Z_2$ must be  the upper row. The rest
 is easy  to complete  since  $S^7/Z_2$ is the real projective space $RP^7$.   We end up   with the
   following  picture
\begin{equation}
\begin{tabular}{lllll}
&  & $Z_{2}$ & $===$ & $Z_{2}$ \\
&  & $\downarrow $ &  & $\downarrow $ \\
$G_{2}$ & $\longrightarrow $ & $Spin(7)$ & $\longrightarrow $ & $S^{7}$ \\
$\parallel $ &  & $\downarrow $ &  & $\downarrow $ \\
$G_{2}$ & $\longrightarrow $ & $SO(7)$ & $\longrightarrow $ & $RP^{7}$%
\end{tabular}.
\end{equation}
From this diagram one can  learn just the lower row, somewhat unexpected, until one sees the middle row.
The lower row is also a remainder that the orthogonal groups have torsion \cite{Borel}. 

In what follows, we  incorporate the $SO(8)$ symmetry of F-theory compactifications in the diagrams.  We will give two   diagrams  connecting $SO(8)$ with special holonomy groups. In terms of CY holonomy groups, we have the following picture

\begin{equation}
\label{seq}
\begin{tabular}{llllll}
$SU(3)$ & $\longrightarrow $ & $SU(4)$ & $=Spin(6)$ & $\longrightarrow $ & $%
S^{7}$ \\
$\quad\downarrow $ &  &  & $\downarrow $ &  & $\parallel $ \\
$\quad Spin(7)$ & $\longrightarrow $ &  & $SO(8)$ & $\longrightarrow $ & $S^{7}$
\\
$\quad\downarrow $ &  &  & $\downarrow $ &  &  \\
$\quad X_{13}$ & === &  & $X_{13}$ &  &
\end{tabular}.
\end{equation}
where $ X_{13}$ is a 13-dimensional homogenous space.\\

The  last   diagram is obtained by  asking  the question how does  $Spin(7)$
 act transitively and  isometrically in the seventh sphere $S^7$.  Indeed,  it must be a
 subgroup of $SO(8)$. What about the quotient (homogeneous space)?.  To answer this question,
   we   start  first   with  the following incomplete cross
\begin{equation}
\begin{tabular}{lllll}
&  & $Spin(7)$ & $\longrightarrow $ & $S^{7}$ \\
&  & $\downarrow $ &  & $\parallel $ \\
$SO(7)$ & $\longrightarrow $ & $SO(8)$ & $\longrightarrow $ & $S^{7}$ \\
&  & $\downarrow $ &  &  \\
&  & $??$ &  &
\end{tabular}%
\end{equation}
and then  try to  finish it. Indeed,  $G_2$ lies inside  both  $Spin(7)$  and  $SO(7)$,
 then it must be their intersection and must appear in the upper left corner. The rest
  of  the diagram   can be obtained easily, and the result is
\begin{equation}
\begin{tabular}{lllll}
$G_{2}$ & $\longrightarrow $ & $Spin(7)$ & $\longrightarrow $ & $S^{7}$ \\
$\downarrow $ &  & $\downarrow $ &  & $\parallel $ \\
$SO(7)$ & $\longrightarrow $ & $SO(8)$ & $\longrightarrow $ & $S^{7}$ \\
$\downarrow $ &  & $\downarrow $ &  &  \\
$RP^{7}$ & === & $RP^{7}$ &  &
\end{tabular}.
\end{equation}
The new result we learn is just the middle column involving th maximal holonomy group of F-theory compactification to four dimensions.

This completes our study of  the relations between holonomy groups which are  suitable for the compactification
 of  superstrings, M, F-theories  respectively.     We have 
  found three triple exact sequences   explaining  some  links between  these  groups. One of
  the nice results that one gets from   the  diagrams is that one can also   see   the possible
  connections between the corresponding geometries. Indeed, from the following sequence of inclusions
\begin{equation}
SU(3)\longrightarrow G_2\longrightarrow S^6,
\end{equation}
one can see that the manifold with $G_2$ holonomy  can  be constructed  in terms of CY
 three folds with the $SU(3)$  holonomy group \cite{Joyce}. The  construction is given by the following seven dimensional orbifold space
\begin{equation}
X_7(G_2) = \frac{CY_3 \times S^1}{Z_2}
\end{equation}
The  Betti numbers of  of $X_7(G_2)$  can  be fixed by the  Hodge numbers of  $CY_3$, which are given by the number of their two and three non trivial cycles.  $Z_2$ acts as the reverse  transformation ($x \to-x$) on the circle $S^1$  and as an involution in the $ CY_3$ space in order to mantain the  $G_2$ structure. The action on the $ CY_3$ is just a simply complex conjugation of its complex coordinates, ($z_i \to {\bar z}_i$).  In this way,  the associative 3-form $\Psi$ of  $X_7(G_2)$  can be  be expressed as 
\begin{equation}
\Psi= \omega \wedge dx + Re (\Omega).
\end{equation}
It is then easy to see that the $G_2$ structure is preserved by the involution, since both the Kahler form $\omega$ of the $CY_3$ and the one-form $dx$ of the circle change sign while the holomophic $Re(\Omega=dz_1 \wedge dz_2\wedge dz_3)$ is invariant. We can suppose the same thing  for the manifold  with  $Spin(7)$  holonomy, it  can be
    constructed either from  manifold with   $G_2$ holonomy  or  CY$_4$. This
     can be easily seen from the subdiagram (\ref{seq}).

\section{Appendix: The octonions}
We recall here some properties of division algebras \cite{Beaz} in relation with special holonomy
 groups and manifolds. Starting with the real numbers $\mathbf R$, the space $R^2$ becomes
  an algebra with  $i \equiv \{0,1 \}$ and $i^2=-1$: we get the complex number $\mathbf
  C$.
  It is a commutative and distributive division algebra. Adding a second unit
  $j$, $j^2=-1$ a third $ij$ is necessary, with $ij=-ji$, and we obtain the
   division algebra of quaternions $\mathbf H$ in $R^4$. It is anticommutative
    but still distributive. Adding another independent unit $k$ to $i$ and $j$,
     with $k^2=-1, ik=-ki, jk=-kj$, we have to complete with $e_7=(ij)k$ to the
      algebra of octonions $\mathbf O$ in $R^8$, with units $1;i,j,k;ij,jk,ki;(ij)k=-i(jk)$.
      It is neither commutative nor associative, but still a division algebra:
      if $o=u_0+ \Sigma_{i=1}^7 u_i e_i$ we have
\begin{equation}
\bar o:=u_0 -\Sigma_{i=1}^7 u_i e_i \qquad {\cal N}(o) = \mbox{norm}(o)\! := \bar o o;\;\;
\mbox{inverse}\, o^{-1}=\frac {\bar o}{{\cal N}(o)}.
\end{equation}

The associator $\{o_1,o_2,o_3 \}:=(o_1 o_2) o_3-o_1 (o_2 o_3)$ is
 completely antisymmetric. The four algebras
${\mathbf R},\;{\mathbf C},\;{\mathbf H},\; {\mathbf O}$ are composition algebras, that is, we have
 ${\cal N}(xy)={\cal N}(x) {\cal N}(y)$. The 
automorphism groups of the algebra are easily seen to be
\be
\mbox{Aut}({\mathbf R})=1,\quad \mbox{Aut}({\mathbf C})=Z_2, \quad \mbox{Aut}({\mathbf H})=SO(3),
\quad \mbox{Aut}({\mathbf O}):= G_2.
\ee

The norm-one elements form, for ${\mathbf R},\;{\mathbf C},\;{\mathbf H},\; {\mathbf O}$, respectively
\be
O(1)=Z_2=S^0; \quad U(1)= SO(2)=S^1; \quad Sp(1)=SU(2)=Spin(3)=S^3; \quad \mbox {and}\, S^7.
\ee

Now $S^7$ has an invertible product structure, in particular is paralellizable,
 but is not a group, because nonassociativity. Let us name jokingly $^\backprime Oct(1)^\prime=S^7$.
 One obtains a {\it bona fide} group by stabilizing $S^7$ by the octonion automorphism
 group $G_2$ \cite{Ramond}. The result is $Spin(7) \approx G_2 (\times S^7$; we shall name $Spin(7):=Oct(1)$, where $(\times$  just means twisted product \cite{{Boya4}}.
We recall now the description of compact Lie groups as finitely twisted products
of odd dimensional spheres (Hopf 1941); for details see \cite{Boya4}. For example
in the quaternion case one gets the sequence
\begin{equation}
Sp(1)=Spin(3)=S^3, \quad Sp(1)^2=Spin(4)=S^3\times S^3,\quad Sp(2)=Spin(5)=S^3 (\times S^{7}.
\end{equation}
There are analogous results  for the  octonions, after $G_2$
stabilization.  The series goes up to dim $3$, but not beyond; this
is due to the  lack of associativity.
  We just write the results, adding the sphere exponents
\begin{eqnarray}
&G_2=SOct(1) \approx S^3 (\times S^{11}; \quad  Spin(8)=Oct(1)^2
\approx S^3 ( \times S^7 ( \times S^7 (\times S^{11} \\
&Spin(9):= Oct(2) \approx S^3 ( \times S^7 ( \times S^{11} (\times S^{15};\quad
F_4:=SOct(3) \approx S^3 ( \times S^{11} ( \times S^{15} (\times S^{23} \nonumber
\end{eqnarray}
where by the prefix "$S$" we mean the unimodular restriction (no $S^7$ factors).
This is similar to $SO$ and $SU$ for $\mathbf R$ and $\mathbf C$ respectively.
 The usefulness of the notation can be seen e.g. in the projective line and plane:
\begin{eqnarray}
&HP^1=S^4=Sp(2)/Sp(1)^2 \; \mbox {corresponds to} \; OP^1=S^8= Spin(9)/Spin(8), \\
&CP^2=S^5/S^1=SU(3)/U(2) \; \mbox {corresponds to} \;
OP^2=SOct(3)/Oct(2)= F_4/Spin(9). \nonumber
\end{eqnarray}
The later is called the Moufang plane (Moufang 1933; to call it the Cayley plane is historically inaccurate).

In any case, this use, $G_2\sim SOct(1)$ etc., is just a notational convention, that we find useful, if carefully employed.
We finish by remarking that little use  has been made so far of the fundamental {\it triality}
 property of the $O(8)$ group and the octonions, namely $Out[Spin(8)]\sim Aut/Inner=S_3$,
 the order three symmetric group. Perhaps in a deeper analysis this triality will show up in particle physics.

The necessary properties we need of the division algebra of the octonions $\mathbf {O}$ are
 described above. Here we recall that $G_2$ is the automorphism group of
  the octonions (as $SO(3)$ is $Aut(\mathbf{H})$ and $Z_2=Aut(\mathbf{C})$ ); the reals
   $ \mathbf {R}$ have not autos, hence the representation $8$ of $G_2$ in the octonions
    split naturally in $8=1+7$. Note that  $G_2$ acts transitively in the $S^6$ sphere of unit imaginary octonions.
     This implies the $6$-sphere acquires a quasi-complex structure (Borel-Serre). The
     sequence  reads as follows
\begin{equation}
\label{su3}
SU(3) \rightarrow  G_2 \rightarrow S^6 \quad (8+6=14)
\end{equation}
where $SU(3)$ acts in the equator $S^5 \in R^6$  as the representation ${\bar3}+3$.
Now the octonionic product preserved by $G_2$, as any
 algebra $( xy=z)$, defines  an invariant  $T^1_2$  tensor  and  the conservation of the norm is like
 preserving a quadratic form.  The $T^1_2$ tensor can be seen  then as a $T^0_3$  tensor. Now the
  alternating property of the octonionic product is equivalent to this $T^0_3$ tensor to  become a 3-form in $R^7$, $ \wedge T^0_3$, which is generic, (i.e., they make up an open set). This implies
\begin{equation}
\mbox{dim}\; G_2= \mbox{dim}\; GL(7,R)-\mbox{dim} \;\wedge T^0_3=49-35=14.
\end{equation}
Besides, the dual form $\wedge T^0_4$ is also invariant, implying  $G_2$ is unimodular, i.e.
lies inside $SO(7)$. The dimension $14$ of this $G_2$ can of course be proved directly \cite{Rosenfeld}.

As with respect to $Spin(7)$, it has a real $8$-dimensional
representation as we said, and hence it acts in $S^7$, indeed
transitively. The little group acts in the $S^6$ equator, and it
 is certainly $G_2$. In fact, there is some sense, as explained above, to call $G_2$ and $Spin(7)$
  respectively $SOct(1)$ and $Oct(1)$.

{\bf Acknowledgments.}   
This work has been supported by CICYT (grant FPA-2006-02315) and DGIID-DGA (grant 2007-E24/2), Spain. 
We thank also the support by  F\'isica de altas energias: Part\'iculas, Cuerdas y Cosmologia, A9335/07. AB would like to thank Departamento de F\'{\i}sica Te\'orica of Zaragoza University for
kind hospitality


\begin{thebibliography} {99}
\bibitem{vafa} C. Vafa, {\em  Lectures on Strings and Dualities}, {\tt  hep-th/9702201}.
\bibitem{witten}
E. Witten,  {\em String theory dynamics in various dimensions},  Nucl. Phys. {\bf B443} (1995)85, {\tt
hep-th/9503124}.
\bibitem{townsend}
 P. K. Townsend,  {\em The eleven-dimensional supermembrane revisited},
 Phys. Lett {\bf B350}
(1995)184, {\tt hep-th/9501068}.
\bibitem{PT}   G. Papadopoulos and P. Townsend, {\em  Compactification of $D=11$ supergravity  on spaces of exceptional holonomy},  Phys. Lett  {\bf B357}
(1995)300-306.
\bibitem{Duff} M.  Duff, {\em The World in Eleven Dimensions},  Ins. of Phys. IOP ( Bristol) 1999.
\bibitem{AW}
B. Acharya, E. Witten,  {\em Chiral Fermions from Manifolds of $G_2$ Holonomy}, {\tt hep-th/0109152}.
\bibitem{adil1} A. Belhaj, {\em M-theory on $G_2$  manifolds and the method of (p,q) brane webs}, J.Phys.
 {\bf A37} (2004)5067-5081.
\bibitem{adil2} A. Belhaj, {\em
 F-theory duals of M-theory on $G_2$  manifolds from mirror symmetry},  J.Phys. {\bf A36} (2003)4191-4206,
   {\tt hep-th/0207208}.
\bibitem{adil4}
A.~Belhaj, M.~P.~Garcia del Moral, A.~Restuccia, A.~Segui and
J.~P.~Veiro, {\em The Supermembrane with Central Charges on a G2
Manifold},   J. Phys. {\bf A42}(2009)325201,   {\tt arXiv:0803.1827 [hep-th]}.
\bibitem{adil3}
A. Belhaj, L.B. Drissi, J. Rasmussen, {\em  On N=1 gauge models from geometric engineering in M-theory},
 Class.Quant.Grav. {\bf 20} (2003) 4973-4982, {\tt hep-th/0304019}.
\bibitem{vafaft}
C. Vafa, {\em Evidence for F-theory}, Nucl. Phys. {\bf B 469} (1996) 403415,
{\tt hep-th/9602022}.
\bibitem{Boya2}  L.J. Boya, {\em Arguments for F-theory}, Mod. Phys. Lett  {\bf A21} (2006) 287-304.
\bibitem{Boya1} L.J. Boya,  {\em Special Holonomy Manifolds in Physics},  Rev. Acad. Cien.  Zaragoza,  {\bf 29}
 (2006) 37-48,  {\tt math-ph/0612002}.
\bibitem{BHV1}
C.  Beasley, J. J. Heckman, C. Vafa, {\em  GUTs and Exceptional Branes in F-theory -I}, {\tt arXiv:0802.3391[hep-th]}.
\bibitem{vafa08} C.  Beasley, J. J. Heckman, C.
Vafa, {\em  GUTs and Exceptional Branes in F-theory - II:
Experimental Predictions}, {\tt arXiv:0806.0102[hep-th]}.
\bibitem{ABBS} R. Ahl Laamara, A. Belhaj, L. J. Boya, A. Segui, {\em  On Local F-theory Geometries and Intersecting D7-branes}, {\tt  arXiv:0902.1161[hep-th]}.
     
\bibitem{ABBMS}
R. Ahl Laamara, A. Belhaj, L. J. Boya, L. Medari, A. Segui, {\em  On F-theory Quiver Models and Kac-Moody Algebras }, {\tt  arXiv:0910.4852 [hep-th]}.
 \bibitem{GS1}   M. Green, J.H. Schwarz and E. Witten, {\em Superstring Theory}, (2 volumes), Cambridge
    University Press (1986). 
\bibitem{sen}
A. Sen,  {\em Heterotic string theory and Calabi Yau manifolds in the Green-Schwarz Formalism},
Nucl. Phys. {\bf B355} (1987) 423.
\bibitem{Oo}
 H. Ooguri and Z. Yin, {\em TASI Lectures on Perturbative String Theories}, {\tt hep-
    th/9612254}.
\bibitem{As} P. Aspinwall, {\em K3 Surfaces and String Duality}, {\tt hep-th/9611137}.
\bibitem{Candelas} P. Candelas {\it et al.}, {\em Vaccum configurations for superstrinsg}, Nucl. Phys. {\bf B258} (1985)46.
\bibitem{witten1}
E. Witten,  {\em New Issues in Manifolds of SU(3) Holonomy}, Nucl. Phys.  {\bf B268} (1986) 79112.
\bibitem{Candelas1}
 P. Candelas {\it et al.}, {\em  A Pair of Calabi-Yau manifolds as an exact soluble superconformal theory}, Nucl. Phys.  {\bf B359} (1991) 21-74.
\bibitem{CJS}
E. Cremmer, B. Julia, J.  Scherk, {\em  Supergravity theory in eleven dimensions}, Phys. Lett. {\bf B 76}(1978)409-412.
\bibitem{BBS} A.  Belhaj, L. J. Boya, A. Segui, {\em  Relation Between Holonomy Groups in Superstrings, M and F-theories}, 
 {\tt arXiv:0806.4265[hep-th]}.

\bibitem{awada83}
 M. Awada  et al,  Phys. Rev. Lett {\bf 50} (1983)294-287.

\bibitem{Joyce}  D. Joyce,  {\em Compact Manifolds with Special Holonomy}, Oxford  U. P. 2000.
\bibitem{Joyce2}D. Joyce, {\em Riemannian  Holonomy Groups and Calibrated  Geometry},   Oxford  U. P. 2007.

\bibitem{KN63} S. Kobayashi  and K. Nomizu, { \em Foundations of Differential  Geometry},
Vol 1, II.8.-J. Wiley, NY.1963.
\bibitem{Boya3} L.J. Boya,  {\em Spinors and Octonions},  {\tt math-ph/0409077}.
\bibitem{Borel} A. Borel, {\em  Topology of Lie groups and characteristic classes},   Bull. Am. Math. Soc {\bf 61} (1955) 397-411.
\bibitem{Beaz} J. Baez,  {\em The octonions}, Bull. Am.  Math Soc. {\bf 39} ( 2002)145-205.

\bibitem{Ramond} R. Ramond, {\em Introduction to Exceptional Lie Groups and Algebras},  Caltech  Preprint CALT-68-577
( december 1976), unpublished.
\bibitem{Boya4}  L. J. Boya, {\em  The Geometry of Compact Lie Groups},  Rep. Math Phys {\bf 30} ( 1991) 149-162.
\bibitem{Rosenfeld} B. Rosenfeld, {\em  Geometry of Lie groups}, Kluwer Academie, Dordrecht 1997, p. 57.



 
\end{thebibliography}
\end{document}